# Unary finite automata vs. arithmetic progressions

**Anthony Widjaja To**

**Abstract** We point out a subtle error in the proof of Chrobak's theorem that every unary NFA can be represented as a union of arithmetic progressions that is at most quadratically large. We propose a correction for this and show how Martinez's polynomial time algorithm, which realizes Chrobak's theorem, can be made correct accordingly. We also show that Martinez's algorithm cannot be improved to have logarithmic space, unless L = NL.

## 1 Introduction

A language is *unary* if it is defined over some unary (i.e. singleton) alphabet. Every unary language *L* can be faithfully represented as a set of natural numbers describing the lengths of the words in *L* (a.k.a. Parikh image of *L*). It is well-known [6] that a unary NFA can be determinized and converted to an exponentially large union of arithmetic progressions. It turns out that this exponential blow-up is avoidable. Chrobak [4] showed a fundamental theorem that every unary NFA has an equivalent one in some normal form, commonly called *Chrobak normal form*, which is at most quadratically large. In fact, NFAs in Chrobak normal form can easily be converted into an equivalent union of arithmetic progressions in linear time. However, Chrobak's original construction is non-algorithmic. Recently, there has been some effort to make Chrobak's construction algorithmic. This has resulted in Martinez's polynomial time algorithm [9,10] and Litow's quasi-polynomial time algorithm [7]. These results have been widely applied (e.g., see [2,5,9,10]).

The goal of this paper is two-fold. First, we would like to point out a subtle error in Chrobak's proof in [4], which does not seem to be immediately obvious how to fix. The gist of the error is the incorrect assumption that in a directed graph with a fixed initial point and a fixed end point, one can traverse the cycles in an arbitrary way without taking into account the dependencies between them (see Example 1 below). Since the correctness proof of Martinez's algorithm relies on the correctness proof of Chrobak's construction (see [10]),

Anthony Widjaja To
School of Informatics, Informatics Forum
10 Crichton Street, Edinburgh, EH8 9AB
United Kingdom
E-mail: anthony.w.to@ed.ac.uk

one can no longer assume that Martinez's algorithm is correct. Furthermore, it turns out that the same error also occurs in Litow's proof of correctness of his algorithm [7]. Secondly, we would like to propose a correction for this error and show how Martinez's algorithm can be made correct accordingly. We will also make a simple observation that one cannot improve Martinez's algorithm to have logarithmic working space unless $\mathtt{L} = \mathtt{NL}$.

*Outline* In Section 2 we pinpoint the location of the gap in Chrobak's proof. We show how to bridge this gap in Section 3. Section 4 shows how Martinez's algorithm can easily be fixed and proves that Martinez's algorithm cannot be improved to have logarithmic space unless $\mathtt{L} = \mathtt{NL}$.

*Notations* We use standard notations from automata theory (see [6]). We shall not worry about which unary alphabet $\Sigma$ to use since any unary language $L$ over $\Sigma$ is completely determined by the lengths of the words in $L$. A unary NFA $\mathscr{A}$ is then a quadruple $(Q, q_0, \delta, F)$, where $Q$ is a set of states, $q_0$ is an initial state, $\delta \subseteq Q \times Q$ is a transition function, and $F$ is a set of final states. For $q \in Q$, we use $succ(q)$ to denote the set $\{q' : (q, q') \in \delta\}$ of all successors of $q$ in $\mathscr{A}$. We use $\mathbb{N}$ for $\{0, 1, 2, \ldots\}$. Given $a, b \in \mathbb{N}$, an *arithmetic progression* with *offset* $a$ and *period* $b$ is the set $a + b\mathbb{N} := \{a + bk : k \in \mathbb{N}\}$. In the sequel, when we deal with computational problems we always *represent numbers in unary*.

## 2 The error

We first state the important lemma that is used in [4]. For a highly readable proof, see also the recent survey [11] on the Frobenius problem.

**Lemma 1 ([3,8])** *Let $0 < a_1 < \ldots < a_k \leq n$ be natural numbers. Then, if $X$ is the set of all $x \in \mathbb{N}$ for which the linear equation $a_1x_1 + \ldots + a_kx_k = x$ is solvable in natural numbers, then*

$$X = S \cup (a + b\mathbb{N})$$

*where $S \subseteq \mathbb{N}$ contains no numbers bigger than $n^2$, and $a$ is the least integer bigger than $n^2$ that is a multiple of $b := \gcd(a_1, \ldots, a_k)$.*

A unary NFA $\mathscr{A} = (Q, q_0, \delta, F)$ is in *Chrobak normal form* if $Q$ can be described as a union of $\{q_0, \ldots, q_m\}$ and the pairwise disjoint sets $C_1, \ldots, C_k$ such that

- for each $1 \leq i \leq k$, $C_i = \{p_{i,0}, p_{i,1}, \ldots, p_{i,j_i-1}\}$ for some integer $j_i > 0$,
- for each $0 \leq i \leq m-1$, $succ(q_i) = \{q_{i+1}\}$,
- $succ(q_m) = \{p_{1,0}, p_{2,0}, \ldots, p_{k,0}\}$, and
- for every $1 \leq i \leq k$ and $0 \leq h \leq j_i - 1$, $succ(p_{i,h}) = \{p_{i,(h+1 \mod j_i)}\}$.

Figure 1 gives an example of an NFA in Chrobak normal form. In other words, an NFA $\mathscr{A}$ is in Chrobak normal form if the structure of $\mathscr{A}$ is a unique path from $q_0$ to the unique state $q_m$ at which $\mathscr{A}$ can make a nondeterministic choice to one of the few disjoint cycles of possibly different periods. Note that there can be more than one final states in $F$. Unary DFAs can be thought of as a subclass of NFAs in Chrobak normal form where $q_m$ has a unique successor state or is a dead end. There is a simple linear-time procedure for converting NFAs in Chrobak normal form to a union of arithmetic progressions. For example, the NFA in Figure 1 is equivalent to the set $(1 + 0\mathbb{N}) \cup (2 + 0\mathbb{N}) \cup (5 + 3\mathbb{N}) \cup (5 + 4\mathbb{N}) \cup (6 + 4\mathbb{N}) \cup (4 + 2\mathbb{N})$. Conversely, as offsets and periods in arithmetic progressions are represented in unary, there is also a trivial linear-time algorithm that converts a union of arithmetic progressions to an NFA in Chrobak normal form. Simply use the largest offset as the length of the unique path, while selecting the final states approriately. Note that applying this procedure on preeceeding example will give us an equivalent NFA that is different from one in Figure 1. Therefore, NFAs in Chrobak normal form and unions of arithmetic progressions are polynomially equivalent representations of unary regular languages. In the sequel, we shall use these two representations interchangeably.

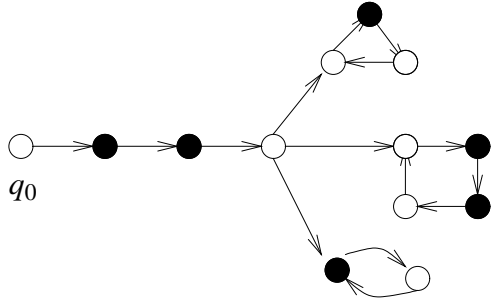

**Fig. 1** An automaton in Chrobak normal form. Filled circles are final states.

**Theorem 1 (Chrobak [4])** *For every n-state unary NFA, there is an equivalent NFA in Chrobak normal form with $O(n^2)$ states, the periods of whose cycles cannot exceed n.*

We now briefly sketch Chrobak's proof in [4] and point out the the subtle error in the proof. In the sequel, a strongly connected component (SCC) is said to be *trivial* if it contains only a single node with no self-loop. An SCC is *nontrivial* if it is not trivial, i.e., either it is a single node with self-loop or it contains at least two nodes. Notice that in a nontrivial SCC for every two nodes $v$ and $v'$ there is always a *nonempty* path from $v$ to $v'$.

*Proof* Suppose that $\mathscr{A} = (Q, q_0, \delta, F)$. If $L(\mathscr{A}) = \emptyset$, then the theorem is obvious and so we shall assume that $L(\mathscr{A}) \neq \emptyset$. Without loss of generality, we may then assume that $F = \{q_F\}$, there is no incoming transition to $q_0$, and that $q_F$ is the unique state with no outgoing transition. Furthermore, we can assume that all states in $Q$ can reach $q_F$ and can be reached from $q_0$. Standard results in graph theory (see [1]) state that we can decompose the directed graph structure of $\mathscr{A}$ into a directed acyclic graph (DAG) $G(\mathscr{A}) = (V, E)$ of SCCs in $\mathscr{A}$ where $V$ and $E$, respectively, denote the vertices and edges of $G(\mathscr{A})$. More precisely, $V$ is the set of all SCCs in $\mathscr{A}$, and that $(D, D') \in E$ iff $\mathscr{A}$ has a transition $(q, q') \in \delta$ for some states $q$ and $q'$ in $D$ and $D'$, respectively. Note that $\{q_0\}$ is the unique node in $G(\mathscr{A})$ with no incoming arc (i.e. "root" of $G(\mathscr{A})$) and $\{q_F\}$ is the unique node in $G(\mathscr{A})$ with no outgoing arc (i.e. "leaf" of $G(\mathscr{A})$). A path $\alpha$ in $G(\mathscr{A})$ from $q_0$ to $q_F$ is called a *superpath* in $\mathscr{A}$. In the sequel, we will also think of a superpath as a subgraph of $\mathscr{A}$ induced by the SCCs in the superpath.

For every superpath $\alpha$ in $\mathscr{A}$, let $L_\alpha$ be the set of all lengths of paths in $\mathscr{A}$ from $q_0$ to $q_F$ that are in $\alpha$. It follows that $L(\mathscr{A}) = \bigcup_\alpha L_\alpha$, where the union ranges over all superpaths in $\mathscr{A}$. Then, it suffices to show that each $L_\alpha$ is a union of some arithmetic progressions $a + b\mathbb{N}$ such that if $a \leq n^2 + n$, then $b = 0$, and if $a > n^2 + n$, then $a < n^2 + 2n$ and $0 < b \leq n$. For then, the number of possible such arithmetic progressions is $O(n^2)$.

We shall not worry about superpaths $\alpha$ consisting only of trivial SCCs as $x \leq n$ for every $x \in L_\alpha$. So, let us assume that $\alpha$ has at least a nontrivial SCC. Let $0 < a_1 < \ldots < a_p \leq n$ be the lengths of all simple cycles in $\alpha$. Let $x \in L_\alpha$ and take a path $R$ in $\alpha$ of length $x$. Then, $x = x_0 + a_1 x_1 + \ldots + a_p x_p$, where $x_0$ is the length of the path obtained from $R$ by deleting cycles. Observe that $x_0 \leq n$. By Lemma 1, it follows that $L_\alpha = L_\alpha^1 \cup L_\alpha^2$, where $L_\alpha^1$ contains no numbers bigger than $n^2 + n$, and $L_\alpha^2$ is an arithmetic progression with period $\gcd(a_1, \ldots, a_n)$ and offset $< n^2 + 2n$. This completes the proof. □

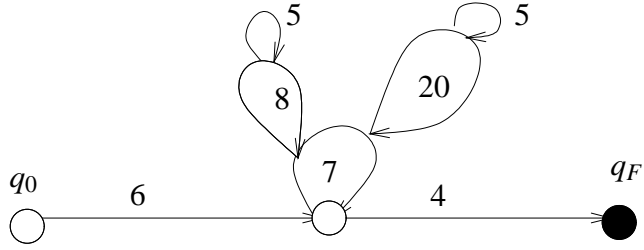

**Fig. 2** A simple example of a superpath with cycle dependencies

Observe that the proof (last paragraph) tacitly assumes that the set $X$ of solutions to the equation $x = x_0 + a_1 x_1 + \ldots + a_p x_p$ corresponds to $L_\alpha$. Although it is the case that $X \supseteq L_\alpha$, it is not the case that $X \subseteq L_\alpha$ in general, which is illustrated in Example 1. This incorrect assumption is also made in the proof of Lemma 5 in [7].

*Example 1* Consider the superpath $\alpha$ given in Figure 2. Here $\alpha$ has only one nontrivial SCC in the middle. Observe that one cannot traverse the cycle of length 5,8, or 20 without traversing the cycle of length 7 at least once. Also, notice that one has to traverse at least once the cycle of length 8 or the cycle of length 20 before we traverse a cycle of length 5. In effect, there are some solutions to the equation $x = 10 + 5x_1 + 7x_2 + 8x_3 + 20x_4$ that are not in $L_\alpha$; for instance, 15, 18, 20, and 30.

## 3 How to fix the proof

In this section, we show how to fill the gap explicated in Example 1. To do this, it suffices to show that each $L_\alpha$ is a union of some arithmetic progressions $a + b\mathbb{N}$ such that if $a \leq 2n^2 + n$, then $b = 0$, and if $a > 2n^2 + n$, then $a < 2n^2 + 2n$ and $0 < b \leq n$. As before, the number of possible such arithmetic progressions is $O(n^2)$. In this section, we call such arithmetic progressions *good*.

Let us take a superpath $\alpha$ with disjoint nontrivial SCCs $D_1, \ldots, D_r$ in some order with $r \geq 1$. Observe that there are only finitely (at most exponentially) many simple paths from $q_0$ to $q_F$ in $\alpha$, say, $\sigma_1, \ldots, \sigma_t$. Define $L_{\sigma_i}$ to be the lengths of all (not necessarily simple) paths from $q_0$ to $q_F$ in $\alpha$ that can be shortened to $\sigma_i$ by deleting cycles. It is clear that $L_\alpha = \bigcup_{i=1}^t L_{\sigma_i}$. So, it suffices to show that $L_{\sigma_i}$ is a union of good arithmetic progressions.

Consider one of the simple paths $\sigma_i$ and suppose that $v_j$ is the point where $\sigma_i$ enters the SCC $D_j$. We shall take *a shortest* (not necessarily simple) cycle $C_j$ at point $v_j$ that visits each node in $D_j$ at least once. The length of $C_j$ cannot exceed $|D_j|(|D_j| - 1)$, where $|D_j|$ is the number of nodes in $D_j$. This is because the length of a shortest path from a node $u$ to another node $u'$ in $D_j$ is at most $|D_j| - 1$. So, we may modify the path $\sigma_i$ as follows: at each $v_j$ before proceeding to the next node in $\sigma_i$, we shall first traverse the loop $C_j$ which will take us back to $v_j$. This is illustrated in Figure 3. Clearly, as $\Sigma_{j=1}^t |D_j| \leq n$, the length of the resulting path $\sigma'_i$ is $|\sigma_i| + \Sigma_{j=1}^{r}|C_j| \leq n + \Sigma_{j=1}^{t}|D_j|^2 \leq n + (\Sigma_{j=1}^{t}|D_j|)^2 = n + n^2$. As $\sigma'_i$ visits each node in $\alpha$ at least once, the set $X$ of of all $x$'s for which $x = |\sigma'_i| + a_1 x_1 + \ldots + a_p x_p$ is solvable in natural numbers is contained in $L_{\sigma_i}$. By Lemma 1, $X = X_1 \cup (g + b\mathbb{N})$, where $X_1$ contains numbers no bigger than $2n^2 + n$, and $g$ is the least integer bigger than $2n^2 + n$ such that $g \equiv |\sigma'_i| \pmod{b}$ where $b = \gcd(a_1, \ldots, a_p)$. Notice that $g < 2n^2 + 2n$. Furthermore, every compound cycle is composed of simple cycles and so we see that $|C_j|$ is a linear combination of $a_1, \ldots, a_p$. This implies that $b$ divides $\Sigma_{j=1}^{t}|C_j|$ and so $g \equiv |\sigma_i| \pmod{b}$.

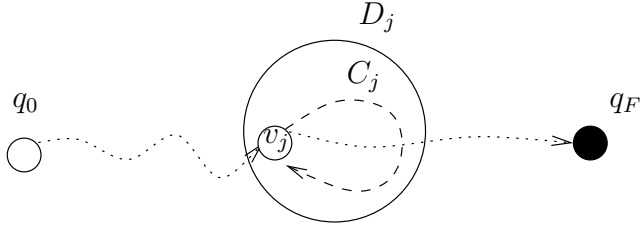

**Fig. 3** The modified path $\sigma'_i$ inside the superpath $\alpha$, zooming in on a particular SCC $D_j$. The dotted line is the original path $\sigma_i$. The dashed line is the additional loop $C_j$.

Conversely, the set $X'$ of $x$'s for which the Diophantine equation $x = |\sigma_i| + a_1 x_1 + \ldots + a_p x_p$ is solvable in natural numbers contains $L_{\sigma_i}$ as we already saw. Therefore, if $m \in L_{\sigma_i}$ and $m > 2n^2 + n$, then Lemma 1 implies that $m \equiv |\sigma_i| \pmod{b}$, where $b = \gcd(a_1, \ldots, a_p)$. We already saw that $g + b\mathbb{N} \subseteq X$ and thus $m \in X$. In summary, we have $L_{\sigma_i} = K \cup X$, where $K$ contains numbers no bigger than $2n^2 + n$. This completes the proof of Theorem 1.

## 4 Martinez's algorithm

**Theorem 2 ([9,10])** *There is a poly-time algorithm converting a given unary NFA to an equivalent one in Chrobak normal form with at most quadratic blow-up.*

Martinez's poly-time algorithm [9,10] works by making sure that each step in Chrobak's proof in [4] can be performed in poly-time. The correctness of Martinez's algorithm relies on the correctness of Chrobak's construction. Due to the subtle error, one has to adjust Martinez's algorithm slightly, which fortunately is not hard to do. Nevertheless, for the sake of completeness, we describe the resulting algorithm. Furthermore, we later show that Martinez's algorithm cannot be improved to have logarithmic working space unless $\mathtt{L} = \mathtt{NL}$.

Martinez's algorithm consists of *four basic steps*. The first step is to ensure that the input NFA $\mathscr{A}$ satisfies $L(\mathscr{A}) \neq \emptyset$ and is in the form required in Chrobak's proof (see above). [If $L(\mathscr{A}) = \emptyset$, just output $\emptyset$.] This can easily be achieved in poly-time by invoking a graph reachability algorithm appropriately several times. So we can write $\mathscr{A} = (Q, q_0, \delta, \{q_F\})$. Second, we apply Kosaraju's poly-time algorithm [1] to decompose $\mathscr{A}$ into the DAG $G(\mathscr{A})$ of SCCs. Third, we compute $\gcd(D)$ for each nontrivial SCC $D$ in $\mathscr{A}$, where $\gcd(D)$ denotes the greatest common divisor of the lengths of all simple cycles in the SCC $D$. Fourth, using this information, we compute the desired union of arithmetic progressions. The only part requiring modification is the fourth step.

Assuming that $G(\mathscr{A})$ has been computed, we shall describe the third step of Martinez's algorithm. Let us view an SCC $D$ as an adjacency matrix $M_D$. By performing simple matrix multiplications, we compute $M_D^1, M_D^2, \ldots, M_D^{|D|}$. We can easily obtain all $j$'s such that the diagonal of $M_D^j$ contains some nonzero entry. Call them $j_1, \ldots, j_r$. These numbers correspond to the lengths of (not necessarily simple) cycles in $D$ of length at most $|D|$. Then, it is the case that $\gcd(j_1, \ldots, j_r) = \gcd(D)$. To see this, first observe that $\gcd(j_1, \ldots, j_r)$ divides $\gcd(D)$ since if $l$ is the length of a simple cycle of $D$, then it has to be one of $j_1, \ldots, j_r$. Conversely, if $j_i$ is the length of a compound cycle in $D$, then $j_i$ can be written as a linear combination of the lengths of simple cycles in $D$. Therefore, $\gcd(D)$ also divides $\gcd(j_1, \ldots, j_r)$. As matrix multiplications and $\gcd(j_1, \ldots, j_r)$ are poly-time computable, we can compute $\gcd(D)$ in poly-time.

Before describing the fourth and final step of Martinez's algorithm, we shall recall an important idea from Chrobak's original proof in [4]. For every nontrivial SCC $D$ in the given NFA $\mathscr{A} = (Q, \delta, q_0, \{q_F\})$, we let $\Pi(D)$ be the set of all superpaths from $q_0$ to $q_F$ whose last nontrivial SCC (i.e. closest to $q_F$) is $D$. Also, let $\Pi_0$ be the set of all superpaths with no nontrivial SCCs. Using the notations in the proof above, we clearly have $L(\mathscr{A}) = \bigcup_{\alpha \in \Pi_0} L_\alpha \cup \bigcup_D \bigcup_{\alpha \in \Pi(D)} L_\alpha$, where $D$ ranges over all nontrivial SCCs in $\mathscr{A}$. As we saw before, $\bigcup_{\alpha \in \Pi_0} L_\alpha$ contains no numbers bigger than $n$. In the following, we use $\gcd(\alpha)$ to denote the greatest common divisor of the lengths of all simple cycles in $\alpha$.

**Lemma 2** *For every nontrivial SCC $D$ in $\mathscr{A}$ with $d := \gcd(D)$, the set $\{x \in \bigcup_{\alpha \in \Pi(D)} L_\alpha : x > 2n^2 + n\}$ is a union of some arithmetic progressions $a + d\mathbb{N}$ such that $2n^2 + n < a < 2n^2 + 3n$.*

*Proof* It suffices to show that, for each $\alpha \in \Pi(D)$ with $g := \gcd(\alpha)$, the set $\{x \in L_\alpha : x > 2n^2 + n\}$ is a union of some arithmetic progressions $a + d\mathbb{N}$ such that $2n^2 + n < a < 2n^2 + 3n$. From the proof of Theorem 1 above, it follows that the set $\{x \in L_\alpha : x > 2n^2 + n\}$ is a union of some arithmetic progressions $a + g\mathbb{N}$ such that $2n^2 + n < a < 2n^2 + 2n$. All simple cycles in $D$ are also simple cycles of $\alpha$ (as $D$ is a subgraph of $\alpha$) and so $g$ divides $d$. Thus we may simply replace each of these $a + g\mathbb{N}$ with a union of the following arithmetic progressions $a + d\mathbb{N}, (a+g) + d\mathbb{N}, (a+2g) + d\mathbb{N}, \ldots, (a+d-g) + d\mathbb{N}$. Since $d \leq n$, each offset $a'$ in any of these arithmetic progression satisfies $2n^2 + n < a' < 2n^2 + 3n$. □

We now describe the final step of Martinez's algorithm. For the sake of simplicity, we give a simplified version of Martinez's algorithm that is is slightly less efficient, albeit still poly-time (see Martinez's papers [9,10] for how to avoid recomputations). As $L(\mathscr{A}) = \bigcup_{\alpha \in \Pi_0} L_\alpha \cup \bigcup_D \bigcup_{\alpha \in \Pi(D)} L_\alpha$ with $D$ ranging over all nontrivial SCCs in $\mathscr{A}$, our strategy is as follows: (i) compute all numbers $i \in L(\mathscr{A})$ that are at most $2n^2 + n$, and (ii) compute a union of arithmetic progressions for each $\bigcup_{\alpha \in \Pi(D)} L_\alpha$ according to Theorem 2. Step (i) is quite easily achieved as we only need to use the standard linear-time algorithm for the membership problem for NFAs. The following lemma will be used in step (ii).

**Lemma 3** *Given a nontrivial SCC $D$ and an arithmetic progression $a + b\mathbb{N}$ with $2n^2 + n < a < 2n^2 + 3n$ and $b = \gcd(D)$, the following statements are equivalent:*

1. $a + b\mathbb{N} \subseteq \{x \in \bigcup_{\alpha \in \Pi(D)} L_\alpha : x > 2n^2 + n\}$.
2. *There exists a path from $q_0$ to $q_F$ of length $a$ that is contained in some superpath $\alpha \in \Pi(D)$.*
3. *There exists a path from $q_0$ to $q_F$ of length $a$ whose last vertex belonging to a nontrivial SCC is in $D$.*

*Proof* Statement (2) and (3) are equivalent by definition. It is obvious that (1) implies (2). Conversely, (2) implies that $a \in S_D$, where $S_D := \{x \in \bigcup_{\alpha \in \Pi(D)} L_\alpha : x > 2n^2 + n\}$. By Lemma 2, there exists an arithmetic progression $a' + b\mathbb{N} \subseteq S_D$ with $2n^2 + n < a' < 2n^2 + 3n$ such that $a \in a' + b\mathbb{N}$. It follows that $a = a' + kb$ for some $k \in \mathbb{N}$ and so $a + b\mathbb{N} \subseteq a' + b\mathbb{N} \subseteq S_D$. □

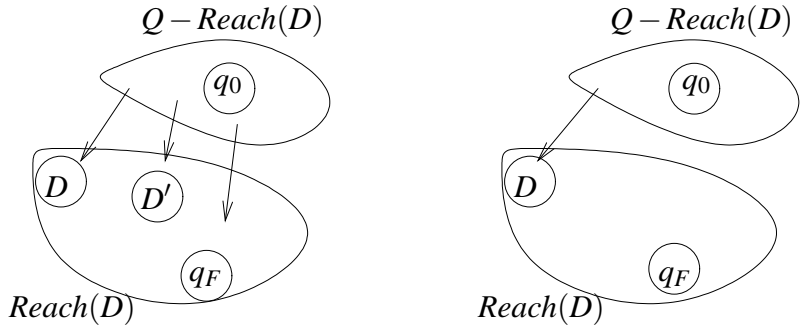

**Fig. 4** The left graph is $\mathscr{A}$ and the right graph is $\mathscr{A}_D$.

By Lemma 3, we shall run through all nontrivial SCC $D$ in $\mathscr{A}$ and collect all arithmetic progressions $a + b\mathbb{N}$ with $2n^2 + n < a < 2n^2 + 3n$ and $b = \gcd(D)$ satisfying statement 3 of Lemma 3. Checking statement 3 of Lemma 3 can be done as follows. Let *Reach*$(D)$ be the set of all states reachable from $D$ in $\mathscr{A}$. Therefore, for each nontrivial SCC $D$, we create a new copy $\mathscr{A}_D$ of $\mathscr{A}$, where we remove all nontrivial SCCs $D'$ with $D' \neq D$ that are reachable from $D$ and all transitions from some state $v \notin Reach(D)$ to a state $v' \in Reach(D) - D$. This is illustrated in Figure 4. The new directed graph $\mathscr{A}_D$ is computable in poly-time by easy applications of graph reachability algorithms. Observe that all paths from $q_0$ to $q_F$ in $\mathscr{A}$ whose last vertex belonging to a nontrivial SCC is in $D$ coincide with all paths from $q_0$ to $q_F$ in $\mathscr{A}_D$. We thus need to check whether there is a path of length $a$ from $q_0$ to $q_F$ in $\mathscr{A}_D$, which is doable in poly-time. Finally, it is easy to check that the entire algorithm runs in poly-time.

We finally remark that a log-space transducer converting a unary NFA to an equivalent one in Chrobak normal form is unlikely to exist. To see this, suppose that such a transducer exists. Observe that there is a simple log-space transducer converting an NFA in Chrobak normal form to an equivalent union of arithmetic progressions. Therefore, combining this with our hypothetical log-space transducer converting a unary NFA to an equivalent one in Chrobak normal form, we have a log-space reduction from the nonemptiness problem for unary NFAs to the nonemptiness problem for unions of arithmetic progressions. The first problem is equivalent under log-space reductions to graph reachability, which is complete for NL, while the second is obviously in L. Hence, we obtain that L = NL. This gives us the following theorem.

**Theorem 3** *There is no log-space transducer converting a unary NFA to an equivalent one in Chrobak normal form, unless* L = NL.

**Acknowledgements** I thank Shunichi Amano, Stefan Göller, and Richard Mayr for proofreading drafts of this paper. I am grateful to Jeffrey Shallit for making Martinez's papers available. The author was supported by EPSRC grant E005039 and ORSAS award.